\documentclass[submission,copyright,creativecommons]{eptcs}

\usepackage{makeidx} 
\usepackage{mathtools}
\usepackage{amsmath}
\usepackage{amssymb}
\usepackage{amsthm}
\usepackage{centernot}
\usepackage{listings}
\usepackage{fancyvrb}
\usepackage{float}
\usepackage{stmaryrd
  ,latexsym,
  fontenc}
\usepackage{semantic}
\usepackage{graphicx}
\usepackage{caption}
\usepackage{xspace}

\usepackage{tikz}
\usetikzlibrary{matrix,shapes,arrows}
\tikzstyle{block} = [rectangle, draw, fill=white!20, text width=4em, text centered, minimum height=3em]
\tikzstyle{line} = [draw, very thick, color=black!90, -latex']

\usepackage{detectEr}
\usepackage{shorthand}
\usepackage{actor}
\usepackage{lang}

\graphicspath{{FIGURES/}}
\newcommand{\qedd}{\hfill\ensuremath{\blacksquare}}

\theoremstyle{definition}

\newtheorem{example}{Example}[section]

\begin{document}

\newcommand{\papertitle}{Improving Runtime Overheads for detectEr}

\title{\papertitle} 
\author{Ian Cassar \masterit
\institute{CS Dept, University of Malta.}
\email{ian.cassar.10@um.edu.mt}
\and
Adrian Francalanza
\institute{CS Dept, University of Malta.}
\email{adrian.francalanza@um.edu.mt}
\and
Simon Said
\institute{CS Dept, University of Malta.}
\email{ssai0008@um.edu.mt}
}
\def\titlerunning{ \papertitle }
\def\authorrunning{A. Francalanza \& I. Cassar}

\maketitle              

\begin{abstract}
We design monitor optimisations for  \detecter, a runtime-verification tool synthesising systems of concurrent monitors from  correctness properties for Erlang programs.  We implement these optimisations as part of the existing tool and show that they yield considerably lower runtime overheads when compared to the unoptimised monitor synthesis.   
\end{abstract}

\section{Introduction}
\label{sec:introduction}

Runtime Verification (RV) \cite{Leu:RV:Overv} is a lightweight verification technique  mitigating the scalability issues associated with  exhaustive verification techniques such as model checking. Low overheads are an important requirement for the viability of any RV framework, where the additional computation 
introduced by the monitors 
should ideally be kept 
to a minimum.

\detecter \cite{CasFra14,FraSey14} is an RV tool for analysing the correctness of concurrent Erlang programs --- the analysis of concurrent programs is 
notoriously hard 
and often leads to state explosion problems.   From a safety correctness property (defined through a formal logic), \detecter\   
generates a system of monitors that execute concurrently with the system under scrutiny, analysing its execution trace, and raising an alert as soon as a violation to the \resp correctness property is detected.   In \cite{FraSey14}, it is shown that the monitors generated by the tool are indeed correct (\eg they only raise an alert when the system violates the \resp property) whereas in \cite{CasFra14} the authors study the relationship between synchronous and asynchronous instrumentation in this setting, establishing (amongst other things) that asynchronous monitoring consistently yields the lowest level of overheads.

In this paper we study optimisation techniques for further lowering the overheads of the tool's asynchronous monitors.\footnote{These optimisations may also be extended to synchronous monitors and should also yield lower overheads to that setting.} The monitor synthesis defined in \cite{FraSey14} uses \emph{concurrent} monitors to parallelise the runtime analysis as much as possible and exploit better the underlying hardware architecture (which nowadays typically includes multiple computing cores). However, in order to simplify the correctness proofs, this synthesis is kept as \emph{regular} as possible: the monitor-generation strategy is the same for every logical construct and does not take into consideration the syntactic context of where that logical construct appears in the  correctness property.       Moreover, the communication organisation of the generated concurrent monitors is also kept \emph{static} throughout the execution of the program, even though certain monitor subsystems become redundant during the runtime analysis.  In this work we address these two potential sources of inefficiency by defining \emph{fine-tuned} organisations of concurrent monitors specifically tailored to different forms of logical formulas;  in addition, these monitors are also able to perform a  degree of \emph{reconfiguration} during the runtime  analysis.     We incorporate the new 
strategies into the existing tool and show that the generated monitors produce lower overheads than the existing monitor translations.

The rest of the paper is structured as follows. \S~\ref{sec:detecter-primer} introduces the tool whereas \S~\ref{sec:and-opt} identifies inefficiencies and proposes solutions. \S~\ref{sec:results} discusses performance improvements and \S~\ref{sec:conclusion} concludes.


\section{\detecter\ Primer}
\label{sec:detecter-primer} 

\paragraph{The Logic:}
\label{sec:detecter-primer-2}

Correctness properties in \detecter\ are expressed through the logic \SHML \cite{aceto:SHML} (a syntactic subset of the modal $\mu$-calculus).   The  \SHML syntax  is defined inductively using the BNF in \figref{fig:logic}.     It is parametrised by a set of boolean expressions, $\bV,\bVV\in\Bool$, equipped with a \emph{decidable} evaluation function, $\bV\Downarrow\vV$ where $\vV \in \sset{\btt,\bff}$, and a set of actions $\actE,\actEE\in\Act$ that may universally quantify over data values. 
It assumes two distinct denumerable sets of \emph{term variables}, $\xV,\xVV,\ldots\in\Vars$ (used in actions  and boolean expressions)  and \emph{formula variables} $\hVarX,\hVarY,\ldots\in\LVars$, used to define recursive (logical) formulas.\footnote{Although we here work up-to $\alpha$-conversion, \detecter\ accordingly renames duplicate variables  during  pre-processing.} Formulas include truth and falsity, \mtru\ and \mfls, conjunctions,  \mand{\hV}{\hVV}, modal necessities, \mnec{\actE}{\hV},  maximal fixpoints to describe recursive properties, \mmax{\hVarX}{\hV}, and conditionals to reason about data, \mif{\bV}{\hV}{\hVV}.  A necessity formula, \mnec{\actE}{\hV}, may contain term variables in \actE\ that \emph{pattern-match} with actual (closed) actions, thus acting as a \emph{binder} for these variables  in the subformula \hV; similarly \mmax{\hVarX}{\hV} is a binder for \hVarX\ in \hV. 

\begin{figure}
  \centering
  \begin{align*}
    \hV,\hVV \in \mFRM & \;\bnfdef\; \mtru \;\bnfsep\; \mfls \;\bnfsep\; \mand{\hV}{\hVV} \;\bnfsep\; \mnec{\actE}{\hV}  \;\bnfsep\; \hVarX \;\bnfsep\; \mmax{\hVarX}{\hV} \;\bnfsep\;  \mif{\,\bV\,}{\,\hV\,}{\,\hVV} 
  \end{align*}
  \begin{align*}
    \hmeaning{\mtru} & \deftxt \Actors \qquad\qquad\qquad\qquad\qquad  \hmeaning{\mfls}  \deftxt \emptyset \qquad\qquad \qquad\qquad\qquad\hmeaning{\mand{\hV}{\hVV}}  \deftxt \hmeaning{\hV} \cap \hmeaning{\hVV}\\
    \hmeaning{ \mnec{\actE}{\hV}} & \deftxt \sset{\actV \;|\;\; \Bigl(\actV \wtraS{\actEE} \actVV  \text{ and } \match(\actE,\actEE) = \sigma\Bigr) \;\text{ implies }\;\actVV \in \hmeaning{\hV\sigma} }\\
    \hmeaning{\mmax{\hVarX}{\hV}} & \deftxt \bigcup \sset{S \;|\; \; S \subseteq \hmeaning{\hV\sset{\hVarX\mapsto S}}} \qquad\qquad\qquad
    \hmeaning{\mif{\,\bV}\,{\,\hV\,}{\,\hVV}}   \deftxt
    \begin{cases}
      \hmeaning{\hV} & \text{if } \bV\Downarrow \btt\\
      \hmeaning{\hVV} & \text{if } \bV\Downarrow \bff
    \end{cases}
  \end{align*}
  \caption{The Logic and its Semantics}
  \label{fig:logic}
\end{figure}

The semantics of the logic is defined for \emph{closed} formulas, over Erlang programs interpreted as a Labelled Transition Systems (LTSs), as shown in \cite{FraSey14}.  In our case, an LTS takes the form $\langle \Actors, \Act\cup\sset{\tau}, \rightarrow \rangle$, where $\actV,\actVV\in\Actors$ 
are nodes denoting actor systems,  $\Act\cup\sset{\tau}$ 
are actions including a silent (internal) action $\tau$, and $\rightarrow$ is a ternary relation of type $\Actors\times(\Act\cup\sset{\tau}) \times \Actors$; we write $\actV \traS{\actE} \actVV$  in lieu of $(\actV,\actE,\actVV) \in \rightarrow$  and use $\actV\wtraS{\actE}\actVV$ to denote $\actV(\traS{\tau})^\ast\cdot\traS{\actE}\cdot(\traS{\tau})^\ast\actVV$.    The semantics is given in Fig.~\ref{fig:logic} and follows that of \cite{FraSey14}. No actor system satisfies  \mfls, whereas all actors satisfying \mtru. Actors satisfying \mand{\hV}{\hVV} must satisfy both \hV\ and \hVV.   Necessity formulas \mnec{\actE}{\hV} are satisfied by all actor systems \actV observing the condition that, \emph{whenever}  pattern-matchable actions   $\actEE$ are performed (yielding substitution $\sigma:: \Vars\rightharpoonup \Val$), the resulting actors \actVV that are transitioned to \emph{must} satisfy $\hV\sigma$. Note that actors that \emph{do not} perform any pattern-matchable actions trivially satisfy \mnec{\actE}{\hV}.  Formula \mmax{\hVarX}{\hV} denotes the maximal fixpoint of the functional $\hmeaning{\hV}$ and allow the logic to be defined over actors with \emph{infinite} behaviour; following standard fixpoint theory \cite{tarski:55}, this is characterised as the union of all post-fixpoints $S\in\pset{\Actors}$ (in Fig.~\ref{fig:logic}, $\sset{\hVarX\mapsto S}$ denotes the substitution of $S$ for $X$).  Finally, a conditional, \mif{\,\bV\,}{\,\hV\,}{\,\hVV}, equates to \hV\ whenever \bV\ evaluates to \btt\  and to \hVV\ when   \bV\ evaluates to \bff.

\begin{example}
  \label{ex:server}  
Consider an Erlang system implementing a \emph{predecessor server} receiving messages of the form \emph{($n$, clientID)} and returning $n-1$ back to \emph{clientID}  whenever $n>0$, but reporting the offending client to an error handler, \eatom{err}, whenever $n=0$.  It may also announce termination of service by sending a message to \eatom{end}. A safety correctness property in our logic would be:
\begin{equation} \label{eq:7:language}
    \mmax{\hVarX}{\mnec{\mrecv{\eatom{srv}}{\etuple{x,y}}}{          
        \left(
          \begin{array}{l}
             \mand{(\mnec{\msend{\eatom{end}}{\_\,}}{\mfls})\;\;\;}{\;\;\;
              (\mnec{\msend{\eatom{err}}{z}}{(\mif{\;(x\neq 0 \vee y \neq z )\;}{\;\mfls\;}{\;\hVarX}) }) 
             }\\
            \quad\mand{}{\;\;
              (\mnec{\msend{y}{z}}{(\mif{\;z=(x-1)\;}{\;\hVarX\;}{\;\mfls} )} )
            }
           \end{array}
        \right)
      }}
\end{equation}
It is a recursive formula, \mmax{\hVarX}{(\ldots)}, stating that, \emph{whenever} the server implementation receives a request (input action), \mnec{\mrecv{\eatom{srv}}{\etuple{x,y}}}{\ldots}, with value $x$ and return (client) address $y$, then it should \emph{not}:
\begin{enumerate}
\item terminate the service (before handing the client request), \mnec{\msend{\eatom{end}}{\_\,}}{\mfls}.
\item report an error, \mnec{\msend{\eatom{err}}{z}}{\ldots}, when $x$ is not $0$, or with a client other than the offending one, $y \neq z$.
\item service the client request, \mnec{\msend{y}{z}}{\ldots}, with a value other than $x-1$.
\end{enumerate}
These conditions are \emph{invariant}; maximal fixpoints capture this invariance for server implementations that \emph{may never terminate} (they are considered correct as long as the conditions above are not violated). \qedd
\end{example}

\paragraph{The Monitor Synthesis:}
\label{sec:monitor-synthesis}

The synthesis algorithm of \cite{FraSey14} aims to be \emph{modular}: it generates independently executing monitor combinators for each logical construct, interacting with one another through message passing.\footnote{Every combinator is implemented as a (lightweight) Erlang process (actor) \cite{Cesarini:2009}, uniquely identifiable by its process ID.  Messages sent to a process are received in its dedicated mailbox, and may be read selectively using (standard) pattern-matching.}  For instance,  the synthesis  \emph{parallelises} the analysis of the subformulas $\hV_1$ and $\hV_2$ in a conjunction $\mand{\hV_1}{\hV_2}$ by $(i)$ synthesising concurrent monitor systems for $\hV_1$ and $\hV_2$ \resp and $(ii)$ creating a conjunction monitor combinator that receives trace events and forks (forwards) them  to the independently-executing monitors of  $\hV_1$ and $\hV_2$.  Since the submonitor systems for $\hV_1$ and $\hV_2$ may be arbitrarily complex (needing to analyse a stream of events before reaching a verdict), the conjunction monitor is \emph{permanent} in the monitor organisation generated, so as to fork and forward event streams of arbitrary length.  It is also worth noting that the synthesis algorithm assumes formulas to be \emph{guarded}, where recursive variables appear under necessity formulas; this is required to generate monitors that implement \emph{lazy} 
unrolling of recursive formulas, thereby minimizing overheads (see \cite{FraSey14} for details).

\begin{figure}[t]
  \centering
  \[\begin{array}{ccc}
  \includegraphics[scale=0.1]{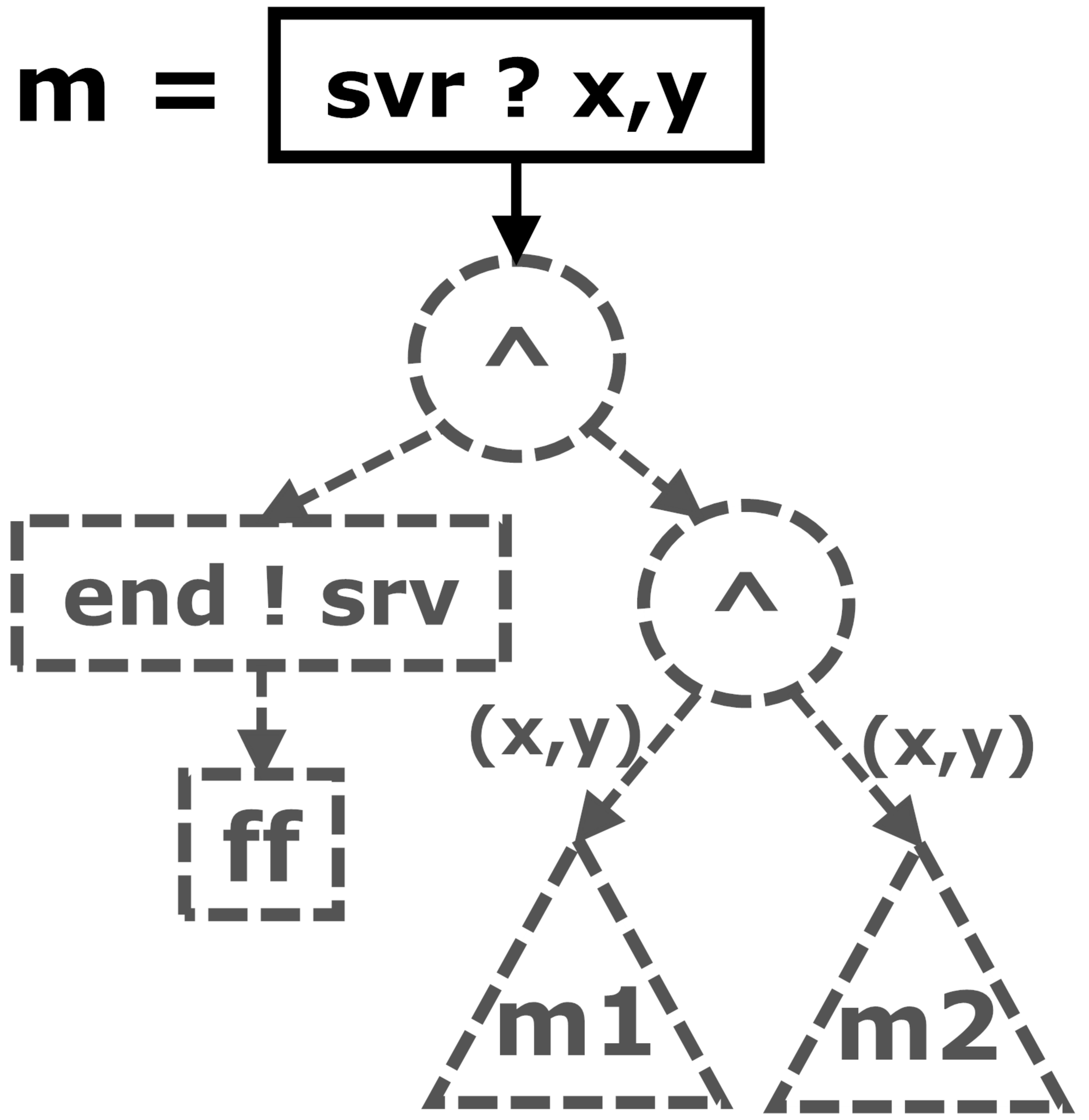} 
&\quad
  \includegraphics[scale=0.1]{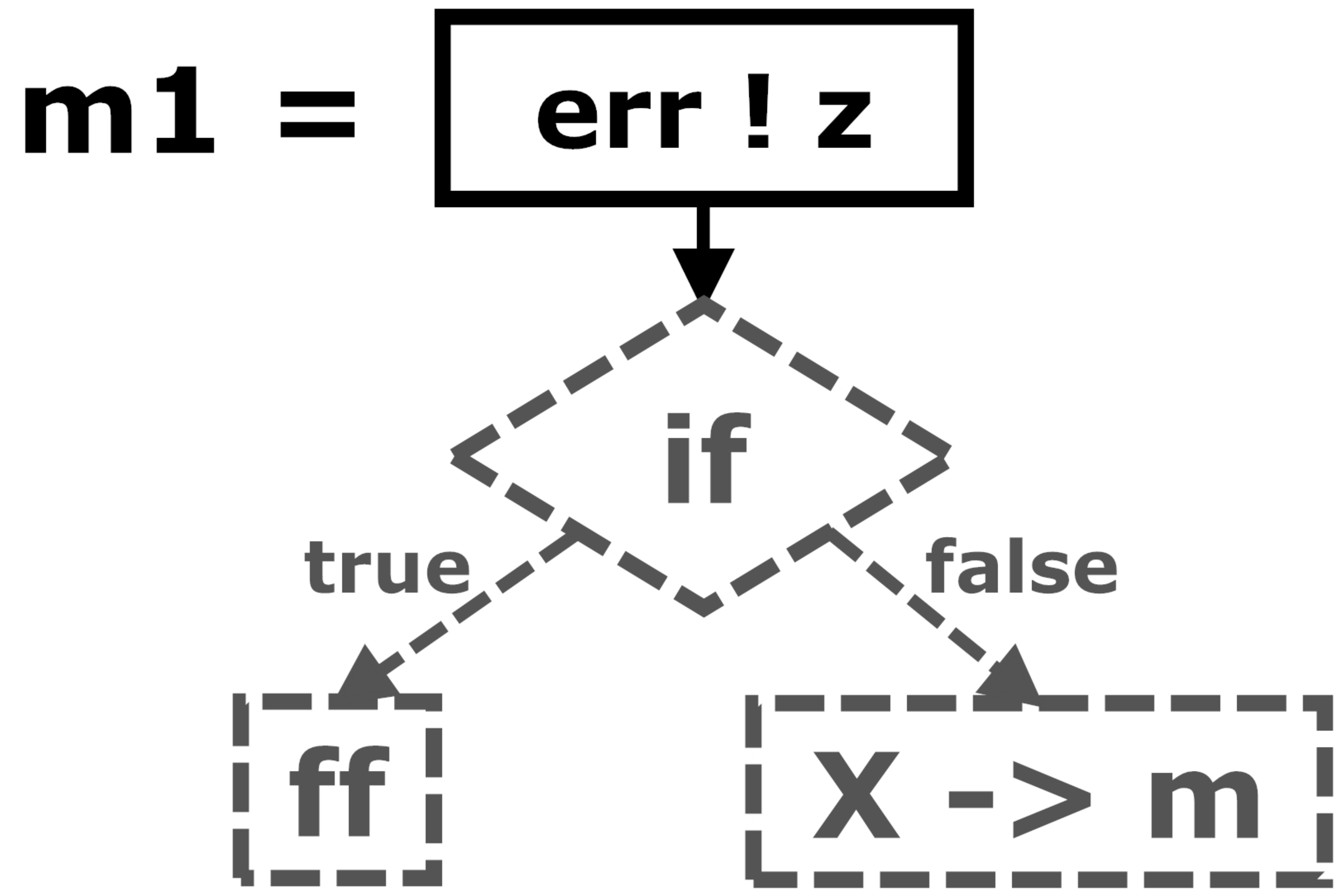}
&\qquad
  \includegraphics[scale=0.1]{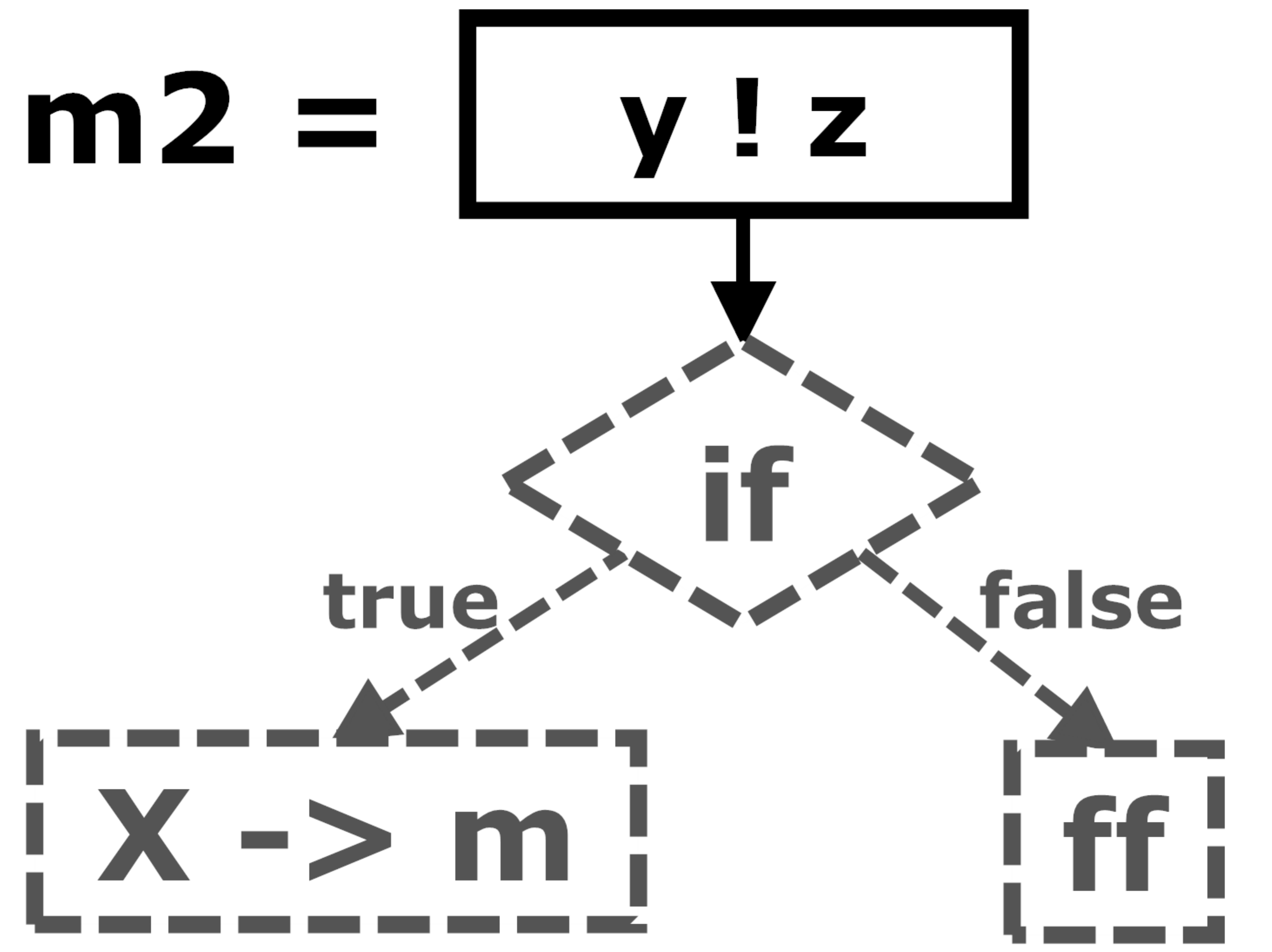}
\end{array}\]
\hrulefill
\[\begin{array}{ccc}
    \begin{array}{l}
\tra{\mrecv{\eatom{srv}}{\etuple{v,c}}}\\
  \qquad\includegraphics[scale=0.1]{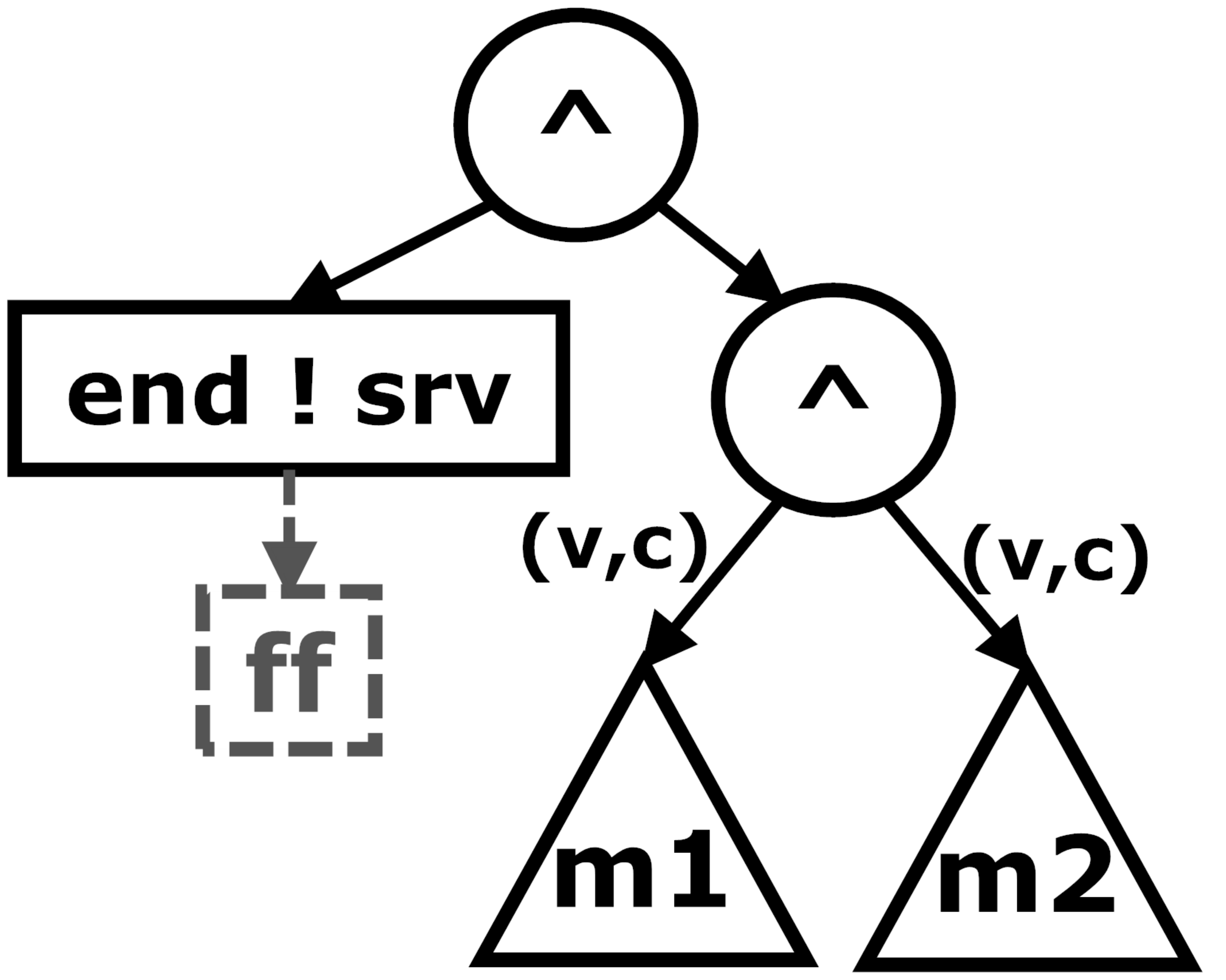}
\end{array}
&\quad
\begin{array}{l}
  \tra{\msend{c}{(v-1)}}\\
  \qquad\includegraphics[scale=0.1]{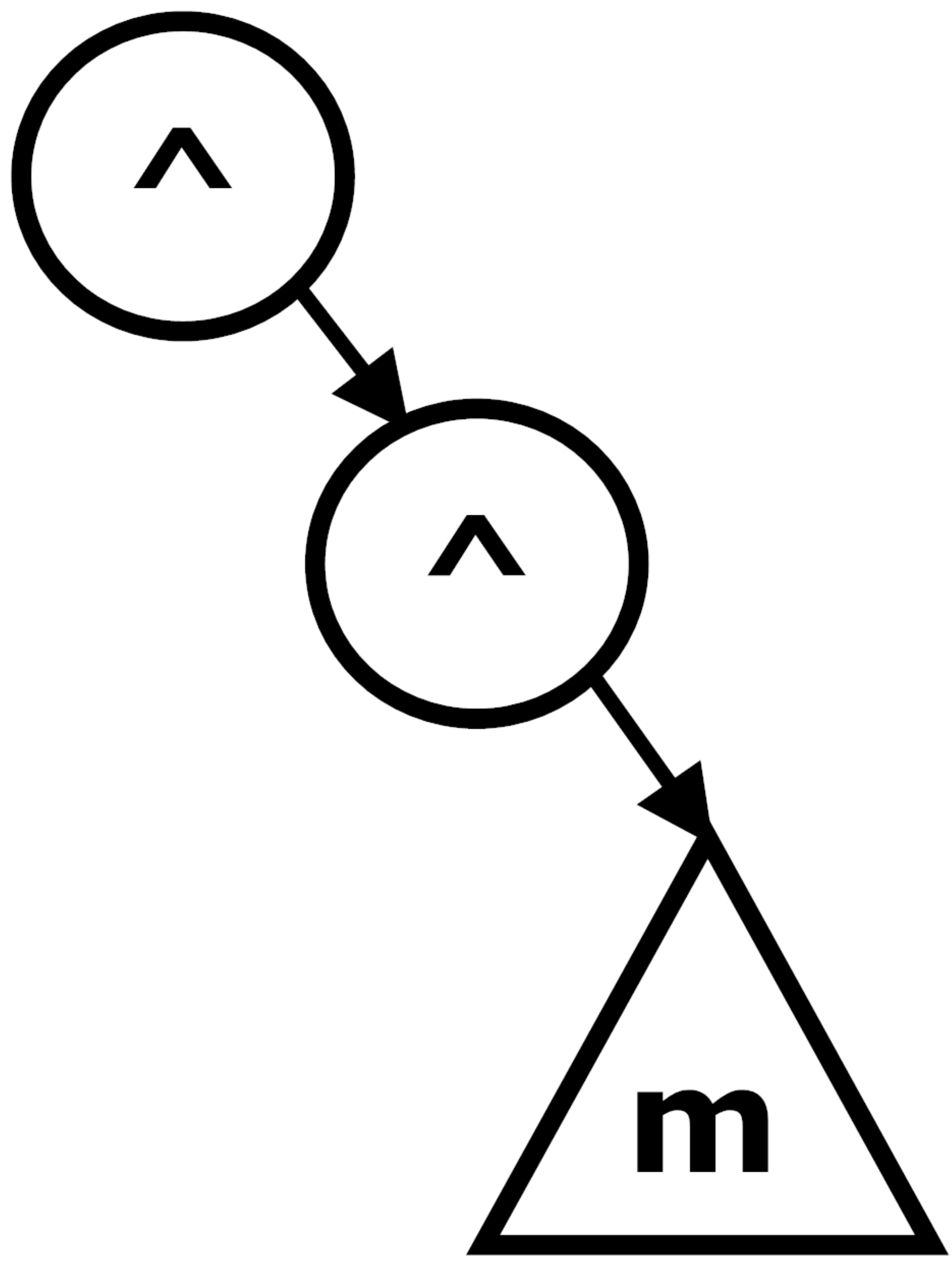}
\end{array}
&\qquad
\begin{array}{l}
    \tra{\mrecv{\eatom{srv}}{\etuple{v',c'}}}\\
  \qquad\includegraphics[scale=0.1]{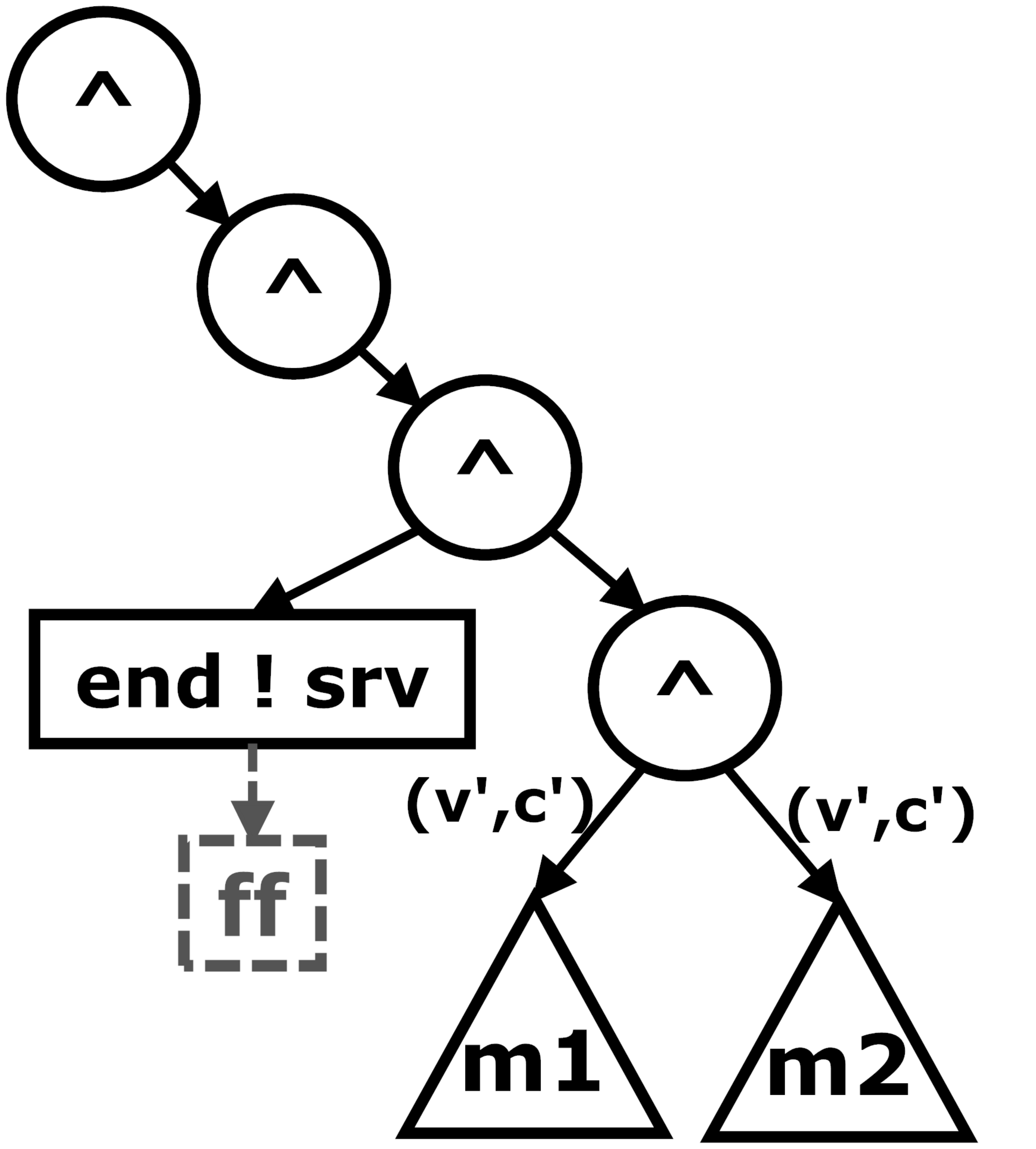}
\end{array}
\end{array}\]
  \caption{Monitor 
    Generation for Formula~(\ref{eq:7:language}) and its execution \wrt trace $\mrecv{\eatom{srv}}{\etuple{v,c}}.\msend{c}{(v-1)}.\mrecv{\eatom{srv}}{\etuple{v',c'}}$}
  \label{fig:monit-gener}
\end{figure}


\begin{example}
  \label{eg:monitor-gen}  From formula~(\ref{eq:7:language}), the monitor organisation $m$ (depicted in Fig.~\ref{fig:monit-gener}) is generated, consisting of one process acting as the combinator for the necessity formula $\mnec{\mrecv{\eatom{srv}}{\etuple{x,y}}}{\hV}$.  If an event of the form $\mrecv{\eatom{srv}}{\etuple{v,c}}$ is received, the process pattern matches it with \mrecv{\eatom{srv}}{\etuple{x,y}} (mapping variables $x$ and $y$ to the values $v$ and $c$ \resp) and  spawns the (dashed) monitor system shown underneath it in Fig.~\ref{fig:monit-gener}, instantiating the variables $x,y$ with $v,c$ \resp  The subsystem consisting of three monitor subsystems, one for each subformula guarded by $\mnec{\mrecv{\eatom{srv}}{\etuple{x,y}}}{}$ in (\ref{eq:7:language}), connected by two conjunction forking monitors.  When the next trace event is received, \eg \msend{c}{v-1}  (a server reply to client $c$ with value $v-1$),  the conjunction monitors replicate and forward this event to the three monitor subtrees. Two of these subtrees do not pattern match this event and terminate; the third subtree (submonitor $m2$) pattern-matches it however, instantiating $z$ for $(v-1)$, evaluating the conditional and unfolding the recursive variable $\hVarX$ to monitor $m$.  If another server request event is received, $\mrecv{\eatom{srv}}{\etuple{v',c'}}$ (with potentially different client and value arguments $c'$ and $v'$), the conjunction monitors forward it to $m$, pattern matching it and generating a subsystem with two further conjunction combinators as before. \qedd     
\end{example}


\section{Optimizations}
\label{sec:and-opt}
Formula~(\ref{eq:7:language}) is a pathological example, highlighting two inefficiencies introduced by the synthesis algorithm of \S~\ref{sec:detecter-primer}.  First, conjunction  monitors mirror closely their syntactic counterpart and can only handle forwarding to \emph{two} sub-monitors.  As a result,   formulas with \emph{nested} conjunctions --- as in formula~(\ref{eq:7:language}) --- translate into organisations of \emph{cascading conjunction monitors} that are inefficient at forwarding trace events.  For instance, the two cascading conjunction monitors of $m$ in Fig.~\ref{fig:monit-gener} replicate the trace event \msend{c}{v-1} \emph{four} times in order to forward it to three sub-monitor systems; the problem is accentuated for higher numbers of nested conjunctions and repeated forwarding.

Second, the current monitor implementation does not perform any monitor reorganisations at runtime.  When a conjunction formula $\mand{\hV_1}{\hV_2}$ is translated, the  conjunction combinator monitor organisation is kept permanent throughout 
its execution because it is assumed that the \resp sub-monitors for $\hV_1$ and $\hV_2$ are long-lived.  This heuristic however does not apply in the case of  formula~(\ref{eq:7:language}), where two out of the three sub-monitors terminate after a single event is received.  This feature, in conjunction with recursive unfolding, creates \emph{chains of indirections} through conjunction monitors with only one child, as shown in Fig.~\ref{fig:monit-gener} (bottom row).

\paragraph{Proposed Solutions:}
\label{sec:proposed-solutions}

\begin{figure}[t]
  \centering
\[\begin{array}{ccc}
    \begin{array}{l}
\tra{\mrecv{\eatom{srv}}{\etuple{v,c}}}\\
  \quad\includegraphics[scale=0.1]{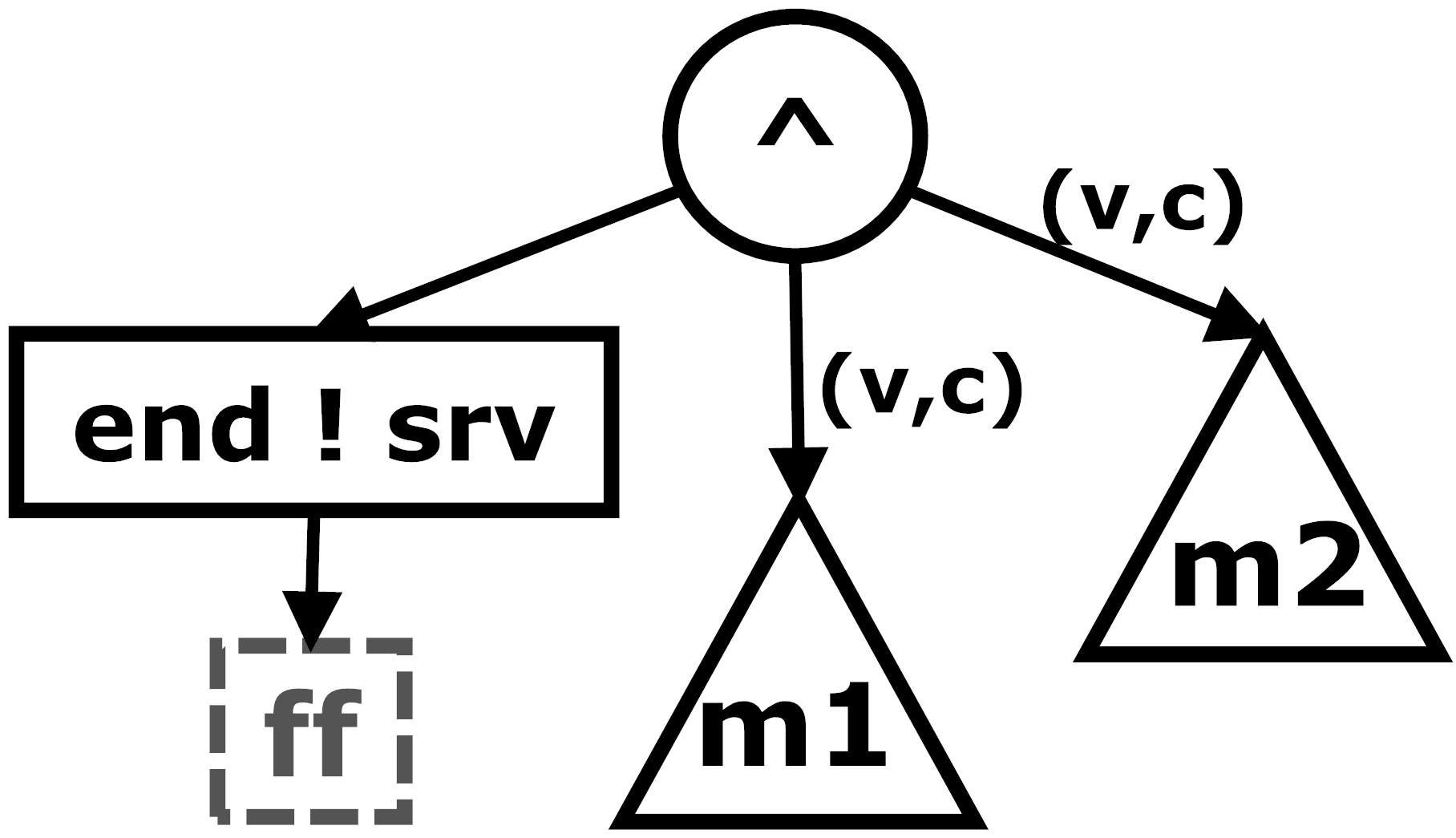}
\end{array}
&\quad
\begin{array}{l}
  \tra{\msend{c}{(v-1)}}\\
  \qquad\includegraphics[scale=0.1]{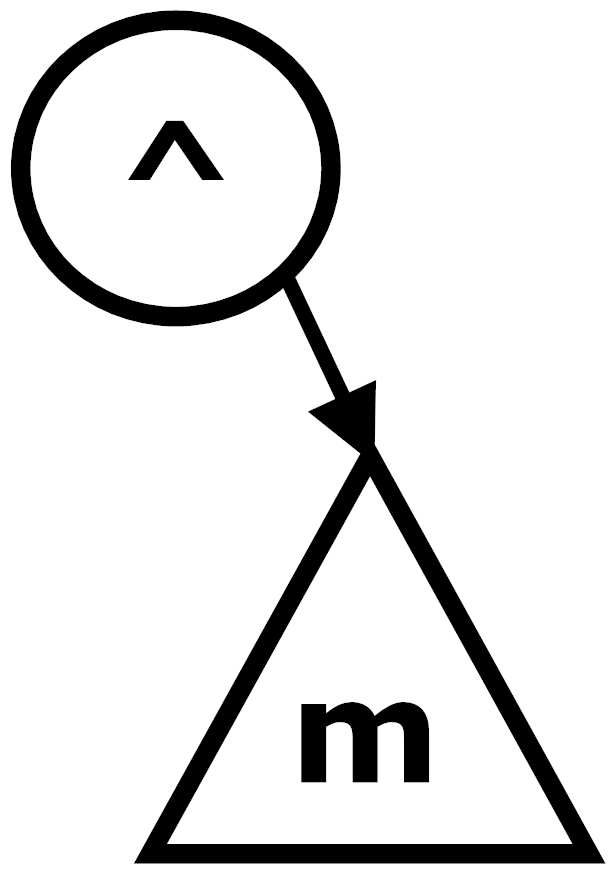}
\end{array}
&\qquad
\begin{array}{l}
    \tra{\mrecv{\eatom{srv}}{\etuple{v',c'}}}\\
  \qquad\includegraphics[scale=0.1]{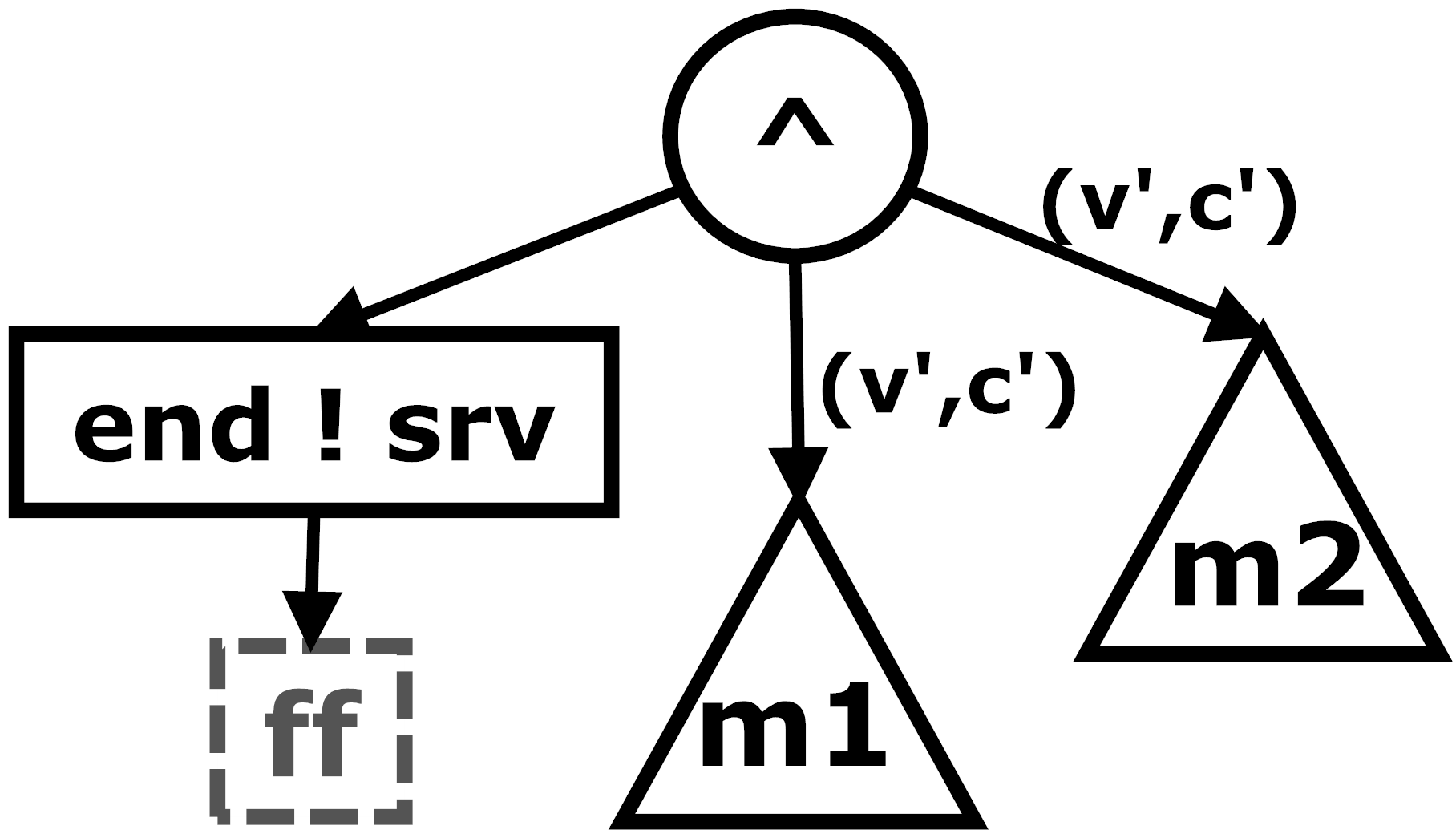}
\end{array}
\end{array}\]
  \caption{Optimised Synthesis for Formula~(\ref{eq:7:language}) and its execution \wrt 
    $\mrecv{\eatom{srv}}{\etuple{v,c}}.\msend{c}{(v-1)}.\mrecv{\eatom{srv}}{\etuple{v',c'}}$}
  \label{fig:monit-gener-opt}
\end{figure}

The first optimisation we introduce is that of conjunction monitor combinators that fork-out to an \emph{arbitrary number} of monitor subsystems. For instance, the corresponding monitor   formula~(\ref{eq:7:language}) would translate into a monitor organisation consisting of \emph{one} conjunction combinator with \emph{three} children (instead of two combinators with two children each)  as shown in Fig.~\ref{fig:monit-gener-opt}(left).      This is more efficient from from the point of view of processes created, but also in terms of the number of replicated messages required to perform the necessary event forwarding to monitor subsystems.  For example, the conjunction combinator of  Fig.~\ref{fig:monit-gener-opt} generates \emph{three} message replications to forward an event to the three sub-monitors  (as opposed to the \emph{four} messages of Fig.~\ref{fig:monit-gener}, as discussed earlier).

The second optimisation considered is that of allowing conjunction monitor combinators to \emph{dynamically reorganise} the monitor configuration  so as to keep the event flow structure as efficient as possible.  In order to keep overheads low, this reconfiguration operation should be kept \emph{local}, where unaffected monitor subsystems should continue with their runtime analysis while the restructuring is in process.  Stated otherwise, the monitor reorganisation happens while  trace events are \emph{still being received}, and the operation needs to guarantee that $(i)$ no trace events are lost $(ii)$ trace events are not reordered.  

Reorganisations are carried out by 
conjunction combinators, 
which are now allowed to \emph{add} and \emph{delete} monitor subsystems from their internal list of children.  For instance, when an event causes a child sub-monitor  to terminate, the parent (conjunction) monitor is sent a termination message which allows it to remove the terminated sub-monitor  from its child-list. 
  
The reconfiguration protocol is kept local (\ie other parts of the monitor graph are unaffected), and is carried by two (multi-child) conjunction combinators in a parent-child  setup.  It proceeds as follows:
\begin{enumerate}
\item When an event causes a child sub-monitor  to become a system with a conjunction combinator at its root, it sends a \emph{merge-request} to its parent. In the meantime the child sub-monitor may start receiving events from its parent and forwards them to its children.
\item When  parent conjunction combinator reads the merge request, it sends a \emph{merge-ack} back to child monitor and  waits for a \emph{merge-msg} from this child; while waiting for this merge message, the parent monitor  stops retrieving further trace events from its mailbox, effectively using it as a buffer for future events that may keep on being received.
\item As soon as the child monitor  receives the \emph{merge-ack}  message, it forwards all the remaining  events in its mailbox to its children. Once it empties its mailbox,  it sends a \emph{merge-msg} back to the parent with a list of its children sub-monitors and waits for a \emph{merge-final} message.
\item Upon receiving \emph{merge-msg} the parent removes the child sub-monitor sending the message and, instead, adds the sub-monitors sent by this child to its own list.  It then sends a \emph{merge-final} to the child monitor and waits for a \emph{merge-complete} message.
\item When the child  receives  \emph{merge-final}, it retrieves any possible merge requests sent by its former children, forwards them in order to its parent, followed by  a \emph{merge-complete} message, and  terminates.
\item When the parent  receives \emph{merge-complete}, it reverts back to its normal operation of trace forwarding.  
\end{enumerate}

The monitor restructuring protocol discussed yields monitor organisations with only \emph{one} (eventual) conjunction node at the root, and a list of monitor subsystems processing the forwarded events (a spider-like configuration).  For instance, for the event trace  $\mrecv{\eatom{srv}}{\etuple{v,c}}.\msend{c}{(v-1)}.\mrecv{\eatom{srv}}{\etuple{v',c'}}$, the synthesised monitor for formula~\eqref{eq:7:language}  yields the evolution shown in Fig.~\ref{fig:monit-gener-opt}.

\section{Evaluation}
 \label{sec:results}
\begin{figure}[!t]
  \centering
\includegraphics[scale=0.6]{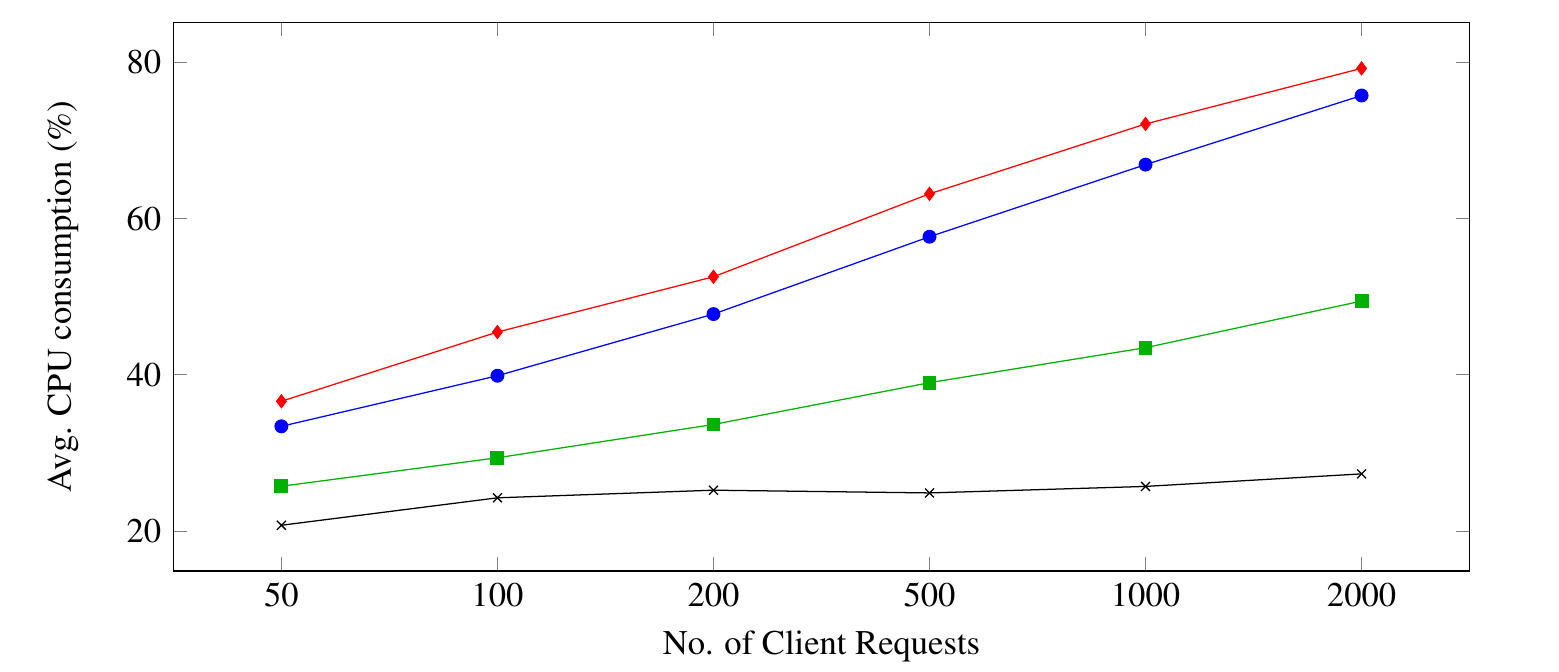}\\
\includegraphics[scale=0.6]{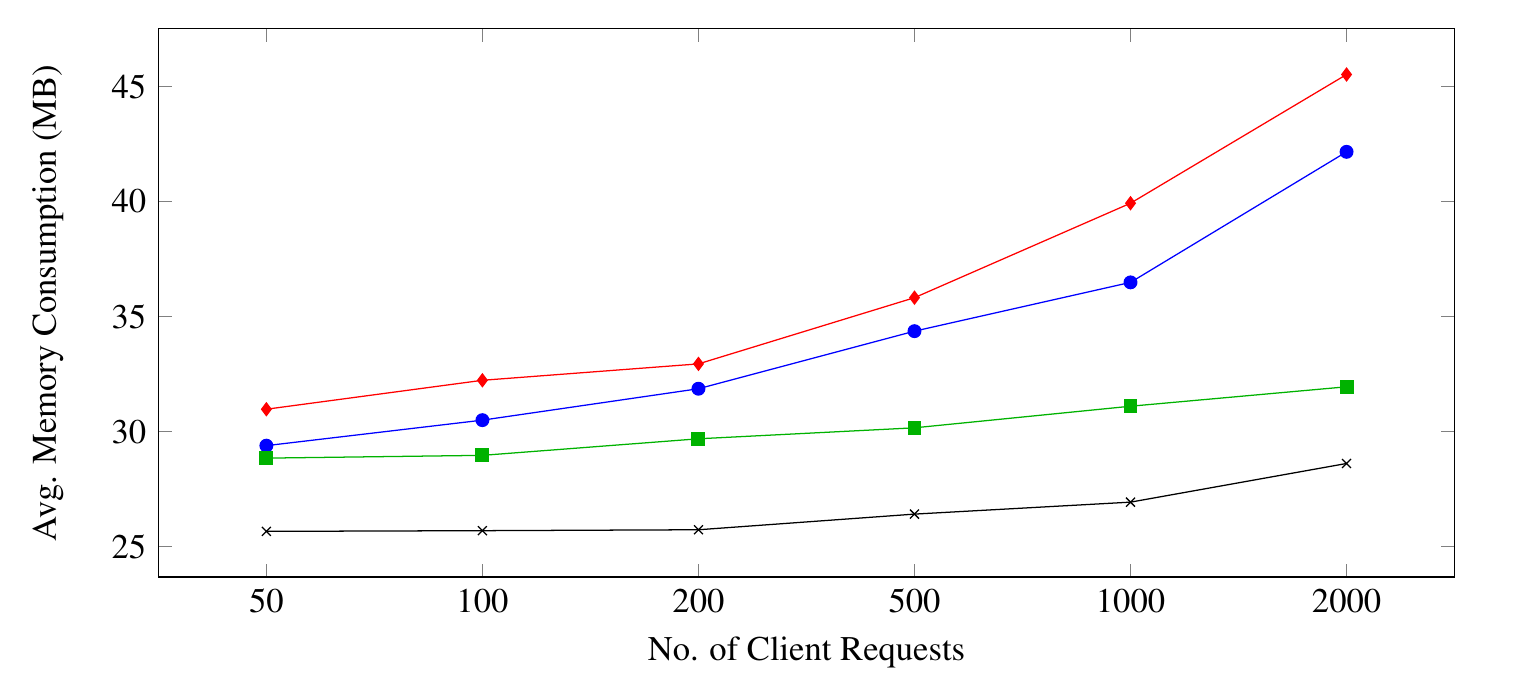}\\
\includegraphics[scale=0.6]{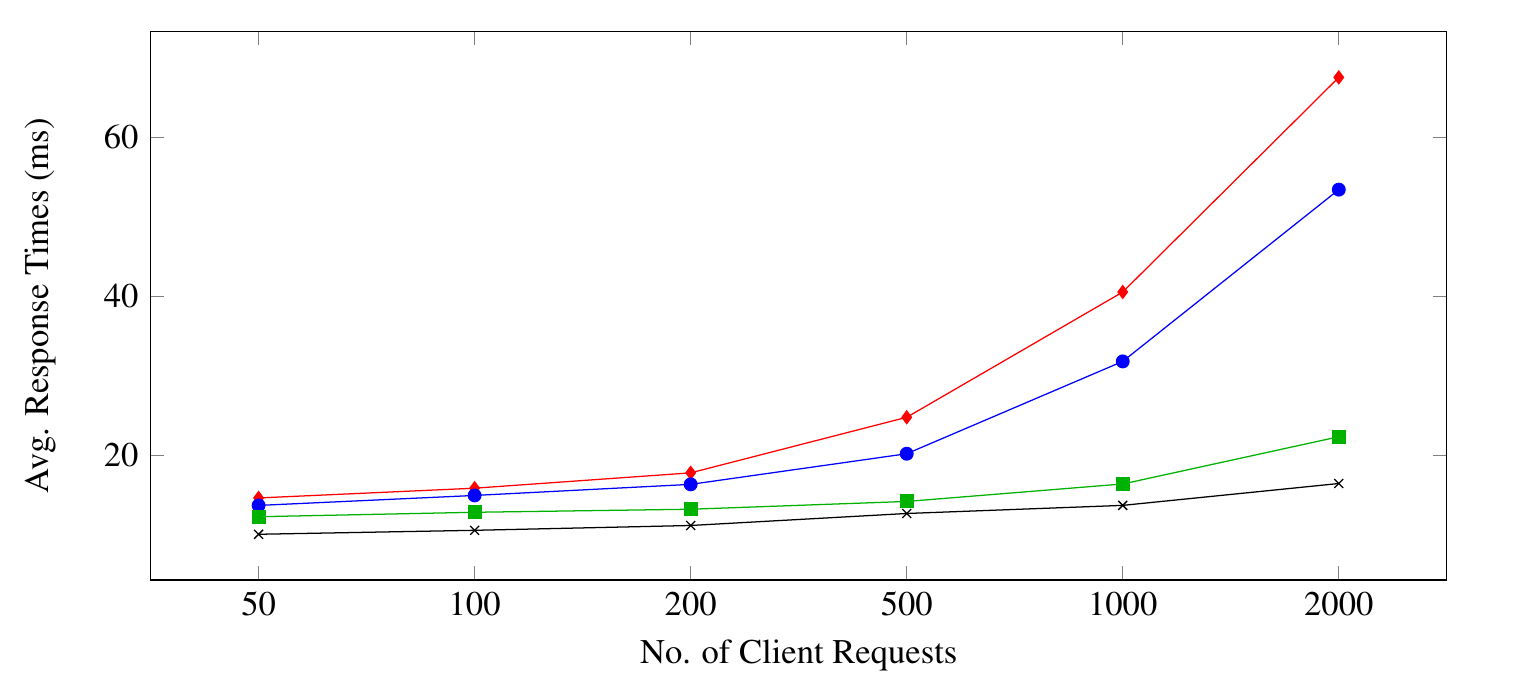}\\
\quad\includegraphics[scale=0.76]{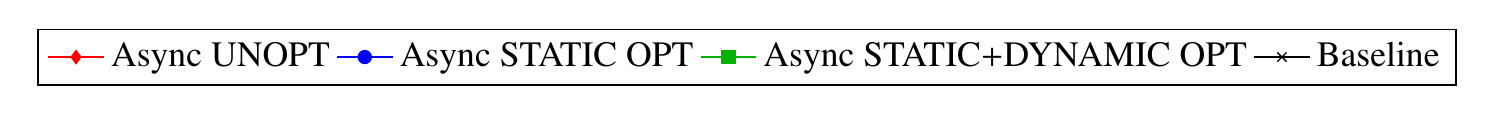}
\caption{Evaluation results}
  \label{fig:opt-eval}
\end{figure}



Through a series of empirical tests, we verify 
whether the monitor optimisations of \S~\ref{sec:and-opt} yield the expected overhead improvements.   In particular, such gains are not obvious for the second optimisation presented in  \S~\ref{sec:and-opt}, where reconfigurations introduce additional computation that may offset the lower overheads obtained from addressing the inefficiencies of redundant monitor code discussed in \S~\ref{sec:detecter-primer}.

The tests are carried out on a third-party commercial software called Yaws \cite{yaws:12}, an HTTP webserver written in Erlang.  In order to keep high levels of throughput, the Yaws server assigns a dedicated (concurrent) handler servicing HTTP requests for every client connection, thereby parallelising processing for multiple clients.   The evaluation is based on a variety of safety properties for the Yaws server implementation, expressed in terms of the logic discussed in \S~\ref{sec:detecter-primer-2}. 
The tests carried out employ three synthesis algorithms to obtain monitors from these properties, namely $(i)$ the unoptimised  synthesis (presented in \cite{FraSey14}), $(ii)$ a monitor synthesis employing multi-child conjunction combinators (without reorganisation) and $(iii)$ a synthesis employing \emph{both} optimisations of \S~\ref{sec:and-opt} (including dynamic reorganisations); these are compared to baseline readings, \ie the unmonitored system.

The tests measure the respective overheads for these three synthesis algorithms, and compare the respective overheads induced \wrt the baseline (unmonitored) Yaws execution for varying client loads.  Overheads are calculated  in terms of $(i)$ the average CPU utilization; $(ii)$ the memory overhead per client request; and $(iii)$ the average time taken for the  server to respond to batches of simultaneous client request.  The experiments are carried out on an Intel Core 2 Duo T6600 processor with 4GB of RAM, running Microsoft Windows 7 and EVM version R16B03.  For each property and each client load, we take the average of three sets of readings.  Since results did not yield substantial variations for the different properties synthesised, we present averaged readings across all properties in the graphs shown in Fig.~\ref{fig:opt-eval}.   

The results show that just using multi-child conjunction combinators yield modest yet consistent gains in terms of CPU usage, memory consumption and average response times, when compared with the two-pronged conjunction combinator of \cite{FraSey14}. However, such monitors still appear to induce non-linear overhead increases, probably created by the chains of monitor indirections created for recursive formula unfolding (see discussion in \S~\ref{sec:and-opt}).  This problem however seems to be rectified for monitors with dynamic reorganisations, as can be seen from  the graphs in   Fig.~\ref{fig:opt-eval}.  In particular, overheads appear to be comparable to the baseline execution for memory consumption.




\section{Conclusion}
\label{sec:conclusion}
\newcommand{\elarva}{e\textsc{Larva}}

We present monitor optimisations for \detecter, an RV tool synthesising concurrent monitors for Erlang correctness properties. We implement these optimisations as part of the existing tool and demonstrate that they yield considerably lower runtime overheads when compared to the original monitor synthesis presented in \cite{FraSey14}.  We conjecture that similar overhead improvements should be observed if these optimisation techniques are applied to the synchronous monitoring studied in \cite{CasFra14}. 

\paragraph{Related Work:}
\label{sec:rel-work}

Several verification and modeling tools \cite{marjan2004,fredlund:mcerlang,
  EfficientDec} for actor-based component systems already exist. Rebeca \cite{marjan2004} is an actor-based modeling language providing automated translation to renowned model checkers like SMV and Promela; timed-rebeca models have also been translated into Erlang.  McErlang \cite{fredlund:mcerlang} is a model-checker specifically targeting Erlang code that uses a superset of our logic; to the best of our knowledge the tool does not consider any verification post-deployment, as in the case of RV.    As far as we are aware, \elarva\ \cite{elarva:2012} is the only other RV tool for Erlang programs.  Similar to the setup studied in this work, it synthesises monitors that use the Erlang Virtual Machine tracing mechanism to obtain system trace events in \emph{asynchronous} fashion.  Apart from the source logic used (\elarva\ properties are described as automata-based specifications), a key difference between this tool and \detecter\ is that \elarva\ produces \emph{monolithic} monitors, as opposed to the component based monitor systems described in this paper; as a result, the optimisation techniques discussed do not apply.    In \cite{EfficientDec}, Sen \etal explore a decentralized (choreographed) monitoring approach as a way to reduce the communication overheads that are usually caused by a centralized approach and implement it in terms of an actor-based tool called \textsc{DiAna}. It would be interesting to explore to what extent the optimisations presented in this work can be extended to the distributed setting of \textsc{DiAna}, and whether these optimisations would yield comparable overhead gains.   Our techniques may also be relevant to lower overheads in other component-based monitor synthesis algorithms such as in \cite{BauFal12} (which has a fixed monitor organisation) or in \cite{dmac} (which supports a level of dynamic reorganisation as updatable distributed tables). Similar investigations could also be carried out for the distributed monitoring approaches studied in \cite{FGP12DistribRV}.



\bibliographystyle{eptcs}
\bibliography{references}

\end{document}